\documentclass[11pt]{article}

\usepackage{amssymb,graphicx,url}
\usepackage[round]{natbib}
\usepackage{rotating}

\usepackage[left=3cm,top=4cm,right=3cm,bottom=2cm,nohead]{geometry}

\newtheorem{definition}{Definition}
\newtheorem{assumption}{Assumption}

\def\Naive{Na\"\i ve}

\newcommand{\widebar}{\overline}
\newcommand{\pitilde}{\tilde{\pi}_i}
\newcommand{\sg}{s_g}
\newcommand{\sX}{s_A}
\newcommand{\NX}{N_A}
\newcommand{\PiXhat}{\widehat{\Pi_A}}
\newcommand{\nX}{n_A}
\newcommand{\nY}{n_B}
\newcommand{\DXbar}{\widebar{D_A}}
\newcommand{\DYbar}{\widebar{D_B}}
\newcommand{\DXbarhat}{\widehat{\widebar{D_A}}}
\newcommand{\DYbarhat}{\widehat{\widebar{D_B}}}
\newcommand{\Dgbarhat}{\widehat{\widebar{D_g}}}

\newcommand{\ADXhat}{\widehat{AD_A}}
\newcommand{\ADYhat}{\widehat{AD_B}}
\newcommand{\ADghat}{\widehat{AD_g}}
\newcommand{\di}{d_i}
\newcommand{\tildedi}{\tilde{d}_i}
\newcommand{\CXYhat}{\widehat{C_{\mbox{\tiny AB}}}}
\newcommand{\CYXhat}{\widehat{C_{\mbox{\tiny BA}}}}
\newcommand{\PX}{P_A}
\newcommand{\PXhat}{\widehat{P_A}}
\newcommand{\PXSHhat}{\widehat{P_A^{\mbox{\tiny SH}}}}
\newcommand{\PXNhat}{\widehat{P_A^{\mbox{\tiny N}}}}
\newcommand{\PXHhat}{\widehat{P_A^{\mbox{\tiny H}}}}
\newcommand{\PXVHhat}{\widehat{P_A^{\mbox{\tiny VH}}}}
\newcommand{\PXSShat}{\widehat{P_A^{\mbox{\tiny SS}}}}
\newcommand{\RCDi}{\mbox{RCD}_i}
\newcommand{\Ehat}{\hat{E}}

\begin{document}


\begin{centering}
\large{\bf The Effect of Differential Recruitment, Non-response and Non-recruitment on Estimators for Respondent-Driven Sampling}\vspace{5mm}

\large{Amber Tomas$^{1}$ and Krista J. Gile$^{2}$\\
$^{1}$Department of Statistics, The University of Oxford\\
$^{2}$Department of Mathematics and Statistics, University of Massachusetts - Amherst}

\end{centering}

\begin{abstract}
Respondent-driven sampling is a widely-used network sampling technique, designed to sample from hard-to-reach populations.
Estimation from the resulting samples is an area of active research, with software available to compute at least four estimators of a population proportion.
Each estimator is claimed to address deficiencies in previous estimators, however those claims are often unsubstantiated.
In this study we provide a simulation-based comparison of five existing estimators, focussing on sampling conditions which a recent estimator is designed to address.
We find no estimator consistently out-performs all others, and highlight sampling conditions in which each is to be preferred.
\end{abstract}

\section{Introduction}
\label{s:Intro}

Respondent-driven sampling (RDS) \citep{H1997} is currently a widely used method for sampling from hidden populations.
A \emph{hidden population} is a population for which it is difficult or impossible to compile a sampling frame, and hence traditional sampling and estimation methods can not be used.
However, it is often still important to collect information on these populations, for example to estimate the prevalence of HIV among injecting drug users in a city.  
In this paper, we compare five estimators based on RDS data, under sampling conditions the estimator presented in \cite{H2007} is designed to address.  
In particular, we focus on the problem of estimating a population proportion in a hidden population with two groups, which we refer to as \emph{infected} and \emph{uninfected}.

The basic method used to select a respondent-driven sample is as follows: 
First, a small number of population members are selected as \emph{seeds}, typically selected from among individuals known to the researcher.
Each seed is given a number of \emph{coupons}, each of which has a unique bar-code, and is asked to pass on the coupons to other people they know within the population.
When an individual has received a coupon, they are asked to report to a study centre where information of interest is collected by the researcher (such information is also collected from the seeds).
A small monetary reward is often offered at this stage to encourage response.
The responders are then themselves given coupons, and are asked to hand them on to others they know within the population, usually only to those who have not yet been recruited.
In this manner, after the initial selection of seeds, the sampling is driven by the respondents.
Those who report to the study centre are known to those who have already been selected, and recruitee-recruiter relationships can be determined from the bar-codes of the coupons.
The information available on which to base an estimate is therefore information collected from the respondents and from the recruitment patterns.
Respondents are usually asked how many people they know within the population of interest.
This provides an estimate of \emph{degree}, as described later.

The original RDS paper \citep{H1997} suggested using the sample proportion as an estimator of population proportion (we refer to this as the ``\Naive'' estimator), and showed that this estimator is unbiased under very strong assumptions about the sampling process.
Subsequent papers \citep{SH2004,VH2008,H2007, G2010} have relaxed some of these assumptions and proposed several alternative estimators.
We will refer to these estimators as the \emph{Salganik-Heckathorn (SH) estimator} \citep{SH2004}, the \emph{Volz-Heckathorn (VH) estimator} \citep{VH2008}, the \emph{Heckathorn (H) estimator} \citep{H2007}, and the \emph{Successive Sampling (SS) estimator} \citep{G2010}.
The SH- estimator has been implemented in the freely-available RDSAT software \citep{RDSAT} for several years, and is widely used by researchers implementing RDS.
The H-estimator has been implemented in the latest version of this software, RDSAT $6.0$ Beta.
There is an urgent need to better understand the properties of the Heckathorn estimator in practice, and in comparison to the VH- and SH-estimators.  
The VH-, SS- and SH- estimators are implemented in the RDS {\bf R} package \citet{rdspackage}.

Although the estimators are easy to implement, the nature of their behaviour is not well understood.
The main reason for this is that several of the assumptions which underpin the theoretical frameworks in which the estimators are derived and analysed are not met in practice.
For example, it is generally assumed that sampling is with replacement or that seeds are selected randomly, whereas in practice these conditions almost never hold.
\citet{GH2009} conducted a simulation study to study the VH- and SH-estimators under realistic sampling scenarios which violated these assumptions.  
Since this work was done the Heckathorn estimator has also been added to the RDSAT software and is already being used in practice.  
This estimator is designed to address violations of sampling assumptions that were not studied in \cite{GH2009}.

In particular, all previous estimators assume that sampled individuals always respond, that they always recruit other individuals when requested, and that they recruit uniformly at random from their acquaintances in the population.
In practice, sampled individuals may not respond (non-response), may not always recruit others (imperfect recruitment effectiveness) and they may preferentially recruit neighbours with particular characteristics (differential recruitment).
In this paper we investigate the effect these three deviations from the sampling assumptions have on the behaviour of the estimators.  
Thus, our comparison study is well-positioned to investigate the contributions of the H- estimator, which claims to adjust for differential recruitment and recruitment effectiveness \citep{H2007}.

In the remainder of this paper we first present the estimators we will compare, with a particular focus on the SH- and H- estimators.
We then investigate in turn, via a simulation study, the effect of differential recruitment, imperfect recruitment effectiveness and non-response on the bias and variance of the estimators.
In all these simulations the SH- and H- estimates were very similar to each other, so we extend the simulation study further to explore the conditions under which these estimates are likely to differ.
Finally, we summarise the results and make some concluding remarks.

\section{Estimators of a Population Proportion}
\label{s:Estimators}

In this section we briefly present the form of the five RDS estimators we consider in the simulation study. 
We use $\pi_i$ to represent the true, unknown probability of selection for individual $i$.
The set of individuals selected in the sample is denoted by $s$, and $s_g$ denotes the intersection of $s$ and the group $g$.
We assume that the response variable is binary, and refer to the levels as infected and uninfected.
$A$ is used to denote the set of all infected individuals, and $B$ to denote the set of uninfected individuals.
The population quantity of interest is the proportion of infected individuals, denoted $P_A$.
Following standard sampling notation, we denote population totals by upper-case letters and sample quantities by lower-case letters.
Hence $\Pi_g = \sum_{i \in g}\pi_i$, for example.
The population size is denoted by $N$, and the sample size by $n$.
We use $\di$ to denote the degree of individual $i$.
The \emph{degree} of an individual is considered to be the number of other individuals in the population to which that individual could potentially pass a coupon.

Most of the estimators described in this paper are functions of Hansen-Hurwitz \citep{HH1943} estimators, or, in the case of the SS estimator, of the closely related Horvitz-Thompson estimators.
The \emph{Hansen-Hurwitz estimator} of a population total $Y$ is unbiased and given by \citep{HH1943}
\begin{equation}
\label{eq:HHest}
\hat{Y} = 
\sum_{i \in s}\pi_i^{-1}y_i.
\end{equation}
The \emph{generalised Hansen-Hurwitz estimator} of a population mean, $\hat{Y}/\hat{N}$, is not unbiased but is asymptotically unbiased.
We refer to any ratio of Hansen-Hurwitz estimators as a generalised Hansen-Hurwitz estimator.

\subsection{\Naive\ Estimator}
\label{ss:N}

The \Naive\ estimate \citep{H1997} is equal to the sample proportion of infected individuals, i.e.
\begin{equation}
\nonumber 
\PXNhat = \frac{\nX}{n}.
\end{equation}
If the true sampling probabilities are given by $\pi_i, i = 1,\ldots,N$, then $\PXNhat$ is a generalised Hansen-Hurwitz estimator of
\begin{equation}
\label{eq:NaiiveEst}
  \frac{\Pi_A}{\Pi_A+\Pi_B}
\end{equation}
(an explanation is given in appendix \ref{app:HHest}).

Equation (\ref{eq:NaiiveEst}) shows that if the sampling probabilities of all individuals are equal, then $\PXNhat$ is a generalised Hansen-Hurwitz estimator of $\NX/N$.
Thus, under very restrictive assumptions, $\PXNhat$ is asymptotically unbiased for $\PX$\footnote{We use the term asymptotically unbiased somewhat reluctantly, as the asymptotics require infinite population size which is often a poor approximation to the scenarios where RDS is used.  However, since asymptotic arguments are used to justify the use of these estimators, we retain them here, noting that it is precisely the failure of assumptions necessary for these asymptotics that necessitates studies such as this one.}.
If the true probabilities of selection for infected individuals are greater than those of uninfected individuals, then it can be seen from (\ref{eq:NaiiveEst}) that $\PXNhat$ will be biased upwards.

\subsection{Volz-Heckathorn Estimator}
\label{ss:VH}

The Volz-Heckathorn (VH-) estimator \citep{VH2008} is given by
\begin{equation}
\label{eq:PXVHhat}
\PXVHhat = \frac{\sum_{i \in s_A}1/\tildedi}{\sum_{i \in s} 1/\tildedi},
\end{equation}
where $\tildedi$ denotes the reported degree of individual $i$.
Using the same approach as in Section \ref{ss:N}, if the true selection probability of individual $i$ is $\pi_i$, $i = 1,\ldots,N$, it can be shown that $\PXVHhat$ is a generalised H-H estimator of
\begin{equation}
\label{eq:VHgenHH}
\frac{\sum_{i \in A}\pi_i/\tildedi}{\sum_{i \in A} \pi_i/\tildedi + \sum_{i \in B} \pi_i/\tildedi}.
\end{equation}
The VH-estimator is asymptotically unbiased for $P_A$ if $\pi_i \propto \tildedi$ for all $i$.
Similarly to the case for $\PXNhat$, it can be seen from expression (\ref{eq:VHgenHH}) that if the true probabilty of selection for infected individuals is larger than assumed, then $\PXVHhat$ is likely to be biased upwards.

\subsection{Successive Sampling Estimator}
\label{ss:SS}

The Successive Sampling (SS-) estimator \citep{G2010} is very similar in form to the VH-estimator.  
The difference is in the estimation of the sampling weights.  
Rather than assuming sampling probabilities proportional to degree as in $\PXVHhat$, this estimator substitutes a function of the degree, based on approximating the sampling process as a successive sampling process.  
This process is without replacement, whereas the \Naive\ and VH- estimators are derived based on the assumption of with-replacement sampling (among others).
Therefore, the key contribution of the SS- estimator over the other estimators is the relaxing of the assumption of with-replacement sampling.
Its form is given by:
\begin{equation}
\label{eq:PXSShat}
\PXSShat = \frac{\sum_{i \in s_A}1/\pitilde}{\sum_{i \in s} 1/\pitilde},
\end{equation}
where $\pitilde$ denotes the estimated sampling probability of individual $i$, estimated according to the successive sampling procedure described in \cite{G2010}.
Again, using the approach in Section \ref{ss:N}, if the true selection probability of individual $i$ is $\pi_i$, $i = 1,\ldots,N$, it can be shown that $\PXSShat$ is a generalised H-T estimator of
\begin{equation}
\label{eq:SSgenHT}
\frac{\sum_{i \in A}\pi_i/\pitilde}{\sum_{i \in A} \pi_i/\pitilde + \sum_{i \in B} \pi_i/\pitilde}.
\end{equation}
Note that here the Horvitz-Thompson estimator \citep{HT1952} is used in place of the Hansen-Hurwitz estimator because of the without-replacement sampling assumption\footnote{Although this estimator has the same mathematical form, the difference is that the sampling probabilities used are without-replacement sampling probabilities, as opposed to the draw-wise selection probabilities used in the Hansen-Hurwitz estimator.}.  
Similarly to the case for $\PXNhat$ and $\PXVHhat$, it can be seen from (\ref{eq:SSgenHT}) that if the true probabilty of selection for infected individuals increases relative to uninfected individuals, $\PXSShat$ will tend to increase. 

This estimator also requires knowledge of the population size, $N$.
The sensitivity of the SS- estimator to the value of $N$ is considered in \citet{G2010}.
For all simulations in this paper we assume that the true value, $N = 1000$, is known, and compute estimates using the {\bf R} package {\tt RDS} \citep{rdspackage}.  

\subsection{Salganik-Heckathorn Estimator}
\label{ss:SH}

The Salganik-Heckathorn (SH-) estimator makes use of the fact that it is possible to measure the proportion of within-group and cross-group recruitments in the sample by keeping track of the barcodes of coupons distributed and returned.

The Salganik-Heckathorn estimator is given by
\begin{equation}
\label{eq:PXSHhat}
\PXSHhat = \frac{\CYXhat}{\CYXhat + \CXYhat\frac{\DYbarhat}{\DXbarhat}}.
\end{equation}
Here $\CXYhat$ denotes the proportion of all individuals recruited by members of group $A$ who are members of group $B$, and can be thought of as an estimate of the probability a randomly selected recruit in group $A$ recruits from group $B$.
$\DXbarhat$ is an estimate of mean degree given by
\begin{equation}
\label{eq:DXbarhat}
\Dgbarhat \stackrel{\mathrm{def}}{=} \frac{n_g}{\sum_{i \in \sg}1/\tildedi},
\end{equation}
and is a H-H estimator of $\sum_{i \in g}\pi_i/\sum_{i \in g} \pi_i/\tildedi$.
Note that as the probability of selection of infected individuals increases relative to uninfected individuals, the ratio $\DYbarhat/\DXbarhat$ will tend to decrease and hence, from (\ref{eq:PXSHhat}), $\PXSHhat$ will tend to increase.

Because of its more complicated form, and the stronger assumptions required to derive it,
the SH- estimator is relatively difficult to study analytically study compared to the \Naive- or VH-estimators.

\subsection{Heckathorn Estimator}
\label{ss:H}

This estimator is an extension of the SH-estimator, and was motivated by the need to control for biases introduced by differential recruitment and recruitment effectiveness \citep{H2007}.
Before introducing the form of the estimator, we need to consider the model used in its derivation: that of a Markov chain on degree groups.
\emph{Degree groups} are formed by partitioning the reported degrees into groups of contiguous degrees.
Then assuming there is only one coupon, perfect recruitment effectiveness and no non-response, the sampling process is treated as a Markov chain.
Here, time is indexed by the wave of the sample, and the state of the chain in a given wave is the degree group of the sampled individual.
The transition probability from group $g$ to $g'$ is the probability that a coupon will be passed from an individual in group $g$ to someone in group $g'$.
Quite strong assumptions about respondent behaviour are required for this model to be appropriate, notably assumptions which are stronger than those required for the sampling process to be modelled as a Markov chain on nodes.

The H- estimator is constructed by using an ``adjusted'' degree estimate $\ADXhat$ in place of the estimates of mean degree in (\ref{eq:PXSHhat}).
The adjusted degree estimate is defined as
\begin{equation}
\label{eq:ADXhat}
\ADXhat = \frac{\sum_{i \in s_A}\RCDi}{\sum_{i \in s_A}\left ( \frac{1}{\tildedi}\RCDi\right )}.
\end{equation}
$\RCDi$ is called the ``recruitment component of degree'' for individual $i$, and is formed based on \emph{degree} groups defined by the reported degree of the respondents.
$\RCDi$ is defined as the ratio of the estimated degree group equilibrium probability and the degree group sample proportion. 
That is,
\begin{equation}
\label{eq:RCDi}
\RCDi = \frac{\Ehat_g}{\frac{n_g}{n}}, \ \mbox{for}\ i\ \mbox{in degree group}\ g,
\end{equation}
where $n_g$ is the number of respondents in degree group $g$ and $\Ehat_g$ is the estimated equilibrium probability of being in degree group $g$.
For all degree groups $g, g'$, the probability of transition for the Markov chain from $g$ to $g'$ is estimated by the proportion of recruitments from $g$ to $g'$, $\widehat{C_{gg'}}$.
The estimated equilibrium probability of being in group $g$, $\Ehat_g$, is then calculated from the matrix of estimated transition probabilities \citep[pg 171]{H2007}.

The Heckathorn estimator uses the adjusted degree estimates in place of the unadjusted degree estimates in the SH-estimator, which gives
\begin{equation}
\label{eq:PXHhat}
\PXHhat = \frac{\ADYhat \CYXhat}{\ADXhat \CXYhat + \ADYhat \CYXhat}.
\end{equation}
If $\RCDi = 1$ for all $i$, then it can be seen from (\ref{eq:ADXhat}) that $\ADghat = \Dgbarhat$ for all $g$, in which case $\PXHhat = \PXSHhat$.

To define the degree groups we use the recommended method of \citet{H2007}, to mimic as closely as possible what we believe to be the default method implemented in RDSAT.
In this method the user specifies a ``mean cell size'' $n_c$, which determines the ``aggregation level'' $AL = \sqrt{n/n_c}$.
The range of degrees is then split into $AL$ groups of approximately equal size.
The default value is $n_c = 12$, which is the value we use for all simulations carried out in later sections.

It should be noted that if all recruits from group $g$ only recruit others from group $g$, the estimated equilibrium probability $\Ehat_g$ will be 1.
This means that $\ADXhat$ will be based only on individuals within the ``absorbing'' degree group, which leads to instability.
However, this case did not arise for any of the simulations presented in this paper.

\subsubsection*{Comments on the Heckathorn Estimator}

$\ADXhat$ can be seen as some kind of adjustment for divergence from equilibrium of the Markov chain on degree groups.
Suppose all the assumptions that are required for the sampling process to be described as a Markov chain on degree groups hold.
Then if the Markov chain started in equilibrium, $n_g/n$ is an unbiased estimator of $E_g$ for all degree groups $g$.
Therefore, from (\ref{eq:RCDi}) it can be seen that it is unlikely for the H- and SH- estimates to differ unless the Markov chain on degree groups does not start in equilibrium. 
This is likely to occur if seeds are chosen disproportionately from individuals of low or high degree, for example, or if only infected individuals are chosen as seeds and the degree distribution of infected and uninfected individuals differ.
Further, we would expect to see larger differences between the H- and SH- estimates if the chain is out-of-equilibrium for a larger proportion of the sample, for example for smaller sample fractions or if convergence is slowed by homophily on degree-group.

The above reasoning is based on the assumption that the sampling process is a Markov chain on degree groups.
This is extremely unlikely to ever occur in practice, but may be a good enough model to inform our investigation.
We comment further on this in Section \ref{s:SHHdiff}.

\subsection{Summary}

All the estimators considered substitute some estimate for the true sampling probabilites into a H-H or H-T estimator: the \Naive\ estimator assumes all $\pi_i$ are equal, the VH- and SH- estimators assume $\pi_i \propto \tildedi$, and the SS- estimator uses a succesive sampling estimate of the $\pi_i$.
Although all the estimators considered in this paper can be shown to be asymptotically unbiased under certain conditions (see \citet{N2009} and \citet{G2010} for details), in practice these conditions will not hold.
Of interest is how the estimators compare when the assumptions are violated.
Differential recruitment, imperfect recruitment effectiveness and non-response are all sampling behaviours which violate the assumptions under which the estimators were derived.
How much these types of sampling behaviour affect the estimates is likely to depend on how great the resulting difference is between the true and estimated sampling probabilties \citep{N2009}.
We investigate this via a simulation study.

\section{Design of the Study}
\label{s:SimDesign}

In this section we present the simulation design used for all simulations presented in this paper.
Because it is not possible to consider all configurations of network and sampling parameters, we concentrate on parameterisations which are likely to approximate real-life conditions.
In particular, the level of homophily, mean degree and the number of seeds were chosen to match the characteristics of the pilot data from the CDC surveillance program \citep{AQ2006,GH2009}. 

The population is modelled as an undirected network with 1000 nodes, where each node represents one individual in the population\footnote{From now on we refer to nodes and individuals interchangeably.}. 
An edge between two nodes $i$ and $i'$ means that there is a non-zero probability that $i$ will recruit $i'$ when given a coupon, and vice versa.
The set of potential recruits of an individual $i$ is therefore given by the neighbours of node $i$, and the number of neighbours is equivalent to the degree of $i$, $d_i$.

For all simulations 200 of the nodes are infected, so $\PX = 0.2$.
The overall mean degree of the network is equal to seven.
There is a moderate amount of \emph{homophily} in all the networks. 
Specifically, we fix the relative probabilities of edges between two infected nodes and between and infected and an uninfected node, such that the former is five times as likely.  
In the case of 20\% infected nodes and no differential activity (defined below), this implies the probability of an uninfected-uninfected edge is twice the probability of an uninfected-infected edge. 
\emph{Differential activity} (DA) is defined as the ratio of the mean degree of infected nodes to the mean degree of uninfected nodes, i.e. $\DXbar/\DYbar$.
This parameter is varied throughout the simulations and takes values in $\{0.5,{\bf 1},1.8\}$.
Parameter values given in bold face denote the default values.  
Networks were simulated using the {\bf R} package {\tt statnet} \citep{statnet, statnetjss}.

Given a population connected by the network, we simulate a respondent-driven sampling process.
In every simulation we draw ten seeds.
Unless otherwise stated, seeds are selected at random with probability proportional to degree from all nodes.
Sampling is done without replacement, and all respondents are given two coupons.
We consider samples of size 200 and 500, but unless there are qualitative differences we only present the simulation results for samples of size 200.
The parameterisation of the sampling process relating to differential recruitment, recruitment effectiveness and non-response will be discussed in more detail in the relevant sections.

For any given value of the population and sampling parameters, the simulations are run as follows:
\begin{enumerate}
\item Simulate 1000 networks.
To allow for comparison, we used the same networks as were used in \citet{GH2009}.
\item For each network, simulate one respondent-driven sample.
We therefore have 1000 samples in total.
\item For each sample, calculate the estimates $\PXNhat$, $\PXSShat$, $\PXVHhat$, $\PXSHhat$ and $\PXHhat$.
\end{enumerate}
The mean of the 1000 estimates from any estimator is a measure of the expected value of that estimator under the given population and sampling parameters.
The variance of the estimates is a measure of the sampling variation of that estimator in addition to how sensitive the estimator is to variations in population structure.

\section{The Effect of Differential Recruitment}
\label{s:DifferentialRecruitment}

Whereas the sample design is determined by the survey organiser, differential recruitment is a property of respondent behaviour which is not possible for the survey organiser to observe directly.
Therefore, it is important to consider how robust the estimators are to the presence of differential recruitment.

In this section we show that the bias of all the estimators is significantly affected by differential recruitment.
The direction of this bias can be explained by considering how differential recruitment affects the probabilities of selection $\pi_i$.

We first define the types of differential recruitment which we will consider in this paper.
\begin{definition}[Differential recruitment] Suppose the proportion of neighbours of node $i$ which belong to group $g$ is equal to $p_{gi}$, for all $g \in G$.  
Then differential recruitment exists if and only if the probability $i$ passes a given coupon to a neighbour from group $g$ is not equal to $p_{gi}$ for any $g \in G$ and any $i$.
\end{definition}
In other words, differential recruitment exists if a respondent is more (or less) likely to pass on a coupon to neighbours of some group than if he were choosing uniformly at random from his neighbours.
We will consider two types of differential recruitment:
\begin{enumerate}
\item Within-group differential recruitment, when nodes preferentially recruit neighbours from within their own group, and
\item Between-group differential recruitment, when all nodes preferentially recruit nodes from a particular group.
\end{enumerate}
For each type of differential recruitment we distinguish between infection-group differential recruitment and degree-group differential recruitment.
It is useful to note that either type of differential recruitment can induce the other if the degree distributions of infected and uninfected nodes are not the same.

We first investigate the influence of within-group differential recruitment in some detail, then consider between-group differential recruitment.

\subsection{Within-group Differential Recruitment}
\label{ss:WithinGrpDR}

We model within-group differential recruitment with the following assumption:
\begin{assumption}
\label{ass:DiffRecruit}
Suppose that node $i$ is a member of group $g$, and is currently recruiting.
Then the probability that node $i$ recruits a neighbour in group $g$ is proportional to $K_g$, and the probability that node $i$ recruits a neighbour from some other group $g' \neq g$ is proportional to 1.
\end{assumption}
This is similar to the two-group model of \citet{GS2009}, except that we allow the strength of differential recruitment to vary by group and do not assume that every individual has a non-zero probability of recruiting any other individual.
Clearly if differential recruitment is greater for individuals in group $g$ than for individuals in group $g'$, i.e. $K_g > K_{g'}$, then the sampling probabilities of group-$g$ individuals will increase relative to those of group-$g'$ individuals.
If differential recruitment exists but is similar for both groups, i.e. $K_g \approx K_{g'}$, then the relative sampling probabilities of nodes in groups $g$ and $g'$ should not change dramatically. 

The effect of within-degree group differential recruitment on the estimators of $\PX$ can therefore be seen to depend on the distribution of degree among infection groups.
If it is the same, then within-degree-group differential recruitment will not induce within-infection group differential recruitment, and hence will not cause additional bias when estimating $\PX$.
If the distribution of degree is not the same among infection groups, for example when there exists differential activity, then within-degree-group differential recruitment will induce within-infection-group differential recruitment.
For example, suppose differential activity is equal to $1.8$, so infected nodes have higher mean degree than uninfected nodes.
If nodes of high-degree prefer to recruit other nodes of high-degree, this will induce differential recruitment by infection group, because infected nodes will preferentially recruit other high-degree infected nodes.

Note that it is not always possible to distinguish from the sample if differential recruitment exists, because its effect on the resulting sampling chain is similar to that of \emph{homophily} - the proportion of within-group edges.
For example, suppose the proportion of recruitments from group $A$ to group $B$, $\CXYhat = 0.5$.
One possibility is that 50\% of the neighbours of group $A$ nodes are in group $B$, and recruitments were made uniformly at random.
Alternatively, it might be the case that 70\% of the neighbours of group $A$ nodes are in group $B$, but that group $A$ nodes preferentially recruited other group $A$ nodes.
Thus, sampling probabilities for a given group will increase with either differential recruitment or homophily.

Based on the properties of the estimators considered in Section \ref{s:Estimators}, we can predict whether the estimates will increase or decrease with increasing within-group differential recruitment.
We would expect that as $K_A$ increases relative to $K_B$, that all of the estimates will also increase.

\subsubsection*{Results}

\begin{figure}[h]
\centering
\includegraphics[angle=270,width=\textwidth]{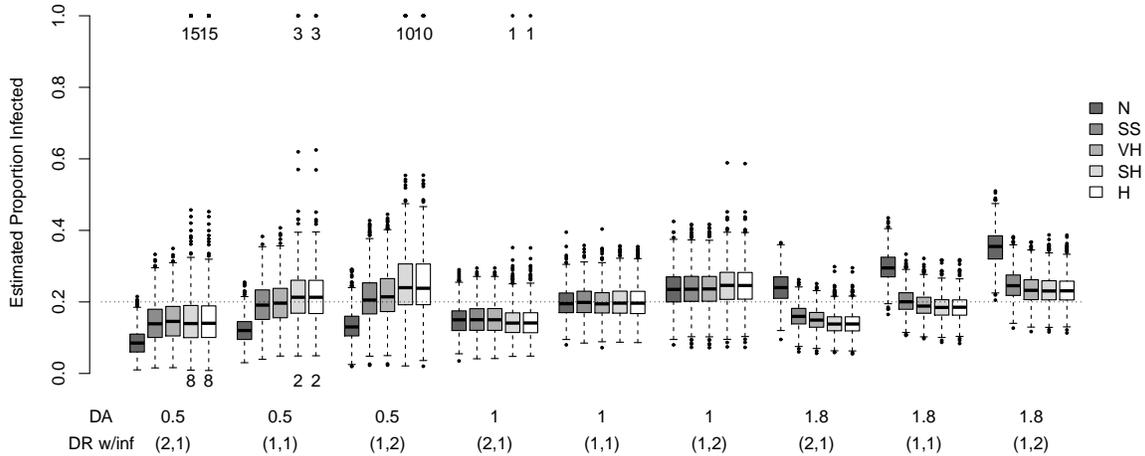}
\caption{Simulation results for varying levels of within-infection-group differential recruitment and differential activity. Differential recruitment is coded as $(K_B,K_A)$.}
\label{fig:DR.dis.w}
\end{figure}

Results of the simulations for three levels of differential activity are shown in Figure \ref{fig:DR.dis.w}.
Note that the counts at the top of the boxplot denote the number of samples for which the estimate was equal to one.
Counts at the bottom of the boxplot denote the number of samples for which an estimate could not be calculated\footnote{The SH- and H- estimators will return a value of 1 if recruitments were made from infection group $A$ to $B$, but not from $B$ to $A$.  In every case that an estimate could not be computed, it was due to there being no recruitments at all from at least one of the infection groups.}.

For all estimators and for all levels of differential activity, as $K_A$ increases relative to $K_B$ the estimates increase, as expected.
It can also be seen that the VH-, SH-, H- and SS- estimates are significantly higher or lower than the sample proportion for differential activity levels of $0.5$ and $1.8$ respectively.
The SS- and VH- estimators are generally less biased than the SH- and H- estimators, and the VH- and SS- estimators also have a lower variance than the SH- and H- estimators.
Other patterns, which are expected from previous studies, are that the SS-, VH- and SH- estimators correct for differential activity (in constrast to the \Naive\ estimator) \citep{GH2009, G2010}, and that the SS- estimates lie between those of the \Naive\ and VH- esitmators \citep{G2010}.
However, there is no evidence that any of the estimators are adjusting for differential recruitment, because the estimates increase or decrease by approximately the same amount as the sample proportion as the differential recruitment level is varied for a given level of differential activity.

There are no significant differences between the H- and SH- estimates for any of the parameter combinations\footnote{All claims of statistical significance are based on paired t-tests of the null-hypothesis of no difference in mean at the 5\% level of significance with Bonferroni correction for multiple comparisons.}.
We also ran simulations with $K_A = K_B \neq 1$ (results not shown), and these verified that the estimates are very similar to the case that $K_A = K_B = 1$, as expected.\\

\begin{figure}
\centering
\includegraphics[width=3in,angle=270]{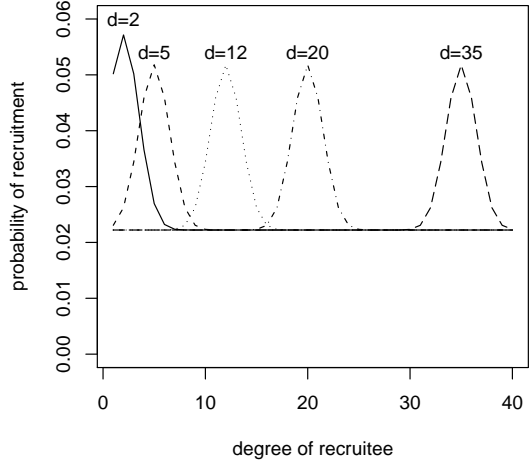}
\caption{Relative probabilities that a node of degree $d$ will recruit a node with degree shown on the x-axis for the simulations with within-degree-group differential recruitment, for several values of $d$.}
\label{fig:WithinDegGrpProbs}
\end{figure}

In order to parameterise within-degree-group differential recruitment for our simulations, we assumed that all nodes preferentially recruit other nodes of similar degree to themselves.
In this way, no matter how the boundaries of the degree groups are defined by the practitioner, within-degree-group differential recruitment will exist.
Figure \ref{fig:WithinDegGrpProbs} shows the probability that a node of degree $i$ will recruit a node of degree $j$, for all $j$ and some $i$, for the simulations with within-degree-group differential recruitment.

\begin{figure}[h]
\centering
\includegraphics[angle=270,width=\textwidth]{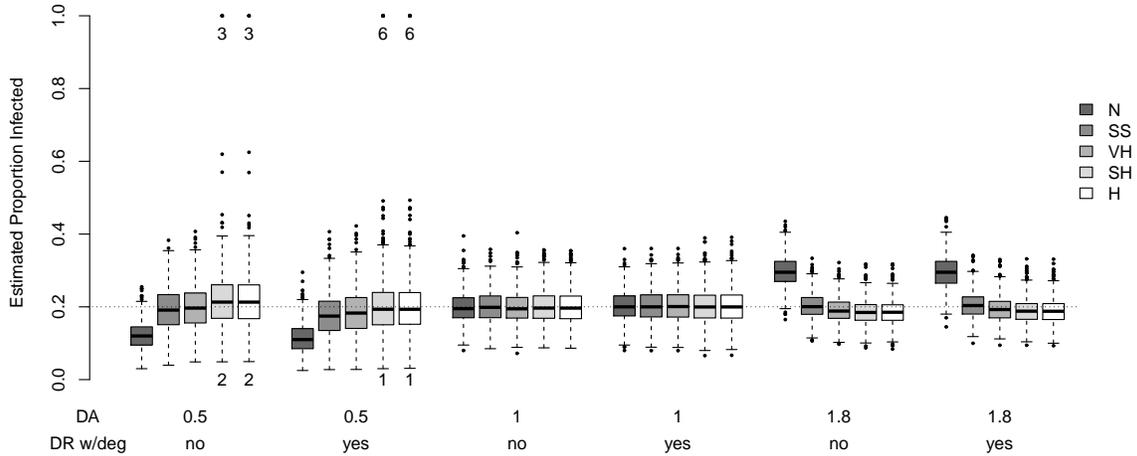}
\caption{Simulation results for varying levels of differential activity, with and without within-degree-group differential recruitment.}
\label{fig:DR.deg.w}
\end{figure}

The simulation results are shown in Figure \ref{fig:DR.deg.w}.
From this figure it can be seen that, for the parameterisation studied in this paper, the effect of within-degree-group differential recruitment on the bias of the estimates is small.
In the absence of differential activity, within-degree group differential recruitment has no significant effect on the bias of the estimators.
When there exists differential activity, then within-degree group differential recruitment reduces (DA = $0.5$) or increases (DA = $1.8$) the mean of each estimator. 
However, the effect of changing levels of differential recruitment is quite small.
This can be explained by the fact that the induced levels of within-infection-group differential recruitment are likely to be similar for both infected and uninfected nodes, i.e. $K_A \approx K_B$.
Hence, from the results of within-infection-group differential recruitment discussed above, the estimators will not show substantial additional bias.

Similarly to the case for within-infection-group differential recruitment, the variance of the SS-estimator is similar to that for the VH-estimator, and the variance of the SH- and H- estimators is slightly greater.
There was only one significant difference between the means of the H- and SH- estimates: when there was both differential recruitment within degree group and differential activity equal to $1.8$ (adjusted p-value $0.0018$).
In this case the mean estimate returned by the H- estimator was $3.7\times 10^{-4}$ greater than the mean for the SH- estimator.

\subsection{Between-group Differential Recruitment}
\label{ss:BetweenGrpDR}

Suppose infected nodes are preferentially recruited by all nodes. 
In this case infected nodes will be more likely to be sampled than if there were no differential recruitment, so the true sampling probability of infected nodes will be higher than that assumed by the estimators.
Therefore, following the arguments of Section \ref{s:Estimators}, we would expect all the estimates to increase.

\begin{figure}[h]
\centering
\includegraphics[angle=270,width=\textwidth]{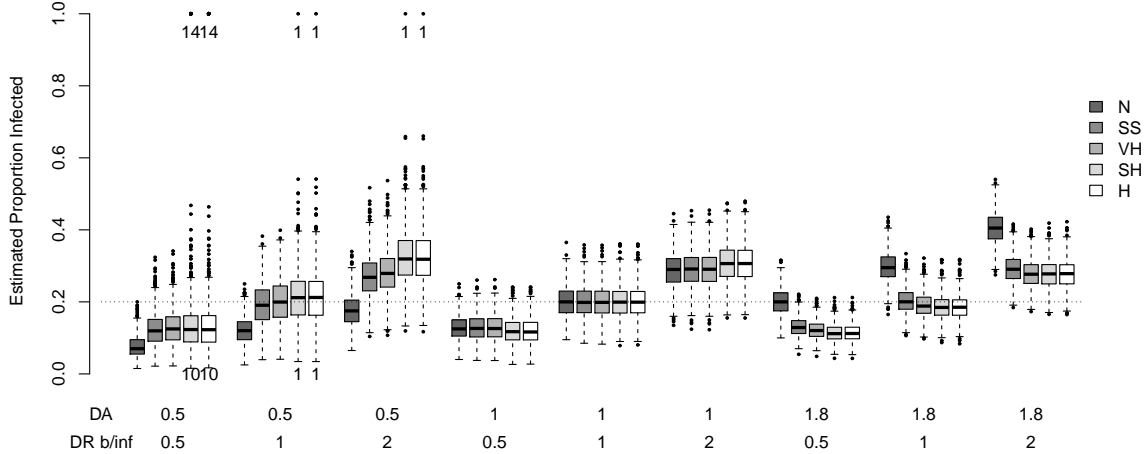}
\caption{Simulation results for varying levels of between-infection-group differential recruitment and differential activity.}
\label{fig:DR.dis.b}
\end{figure}

For our simulations, we parameterised differential recruitment between infection-group by the ratio of the probability a node will recruit an infected node relative to uninfected node.  
Possible values are $\{2,{\bf 1},0.5\}$.
Results of the simulations for three levels of differential activity are shown in Figure \ref{fig:DR.dis.b}.
This shows that for any fixed level of differential activity, as the probability an infected node will be recruited increases relative to the probability an uninfected node will be recruited, all of the estimates of $\PX$ also increase.
There is no evidence that any of the estimators considered correct for differential recruitment in this case, although the SS- and VH- estimators tend to be slightly less biased than the SH- and H- estimators.
The mean of the estimates changes by about the same amount as for the sample proportion as the level of differential recruitment between infection-group is changed.\\

Now consider between-degree-group differential recruitment.
If high-degree nodes are preferentially recruited by all nodes then this is likely to influence the estimates of $\PX$ only if the occurrence of infection is greater or less in nodes of high-degree than in the general population.
That is, we would expect between-degree-group differential recruitment to affect the estimates of $\PX$ only if the degree distribution of infected nodes is different from the degree distribution of uninfected nodes, for example if there exists differential activity.
In this case between-degree-group differential recruitment will induce between-infection-group differential recruitment, and we would expect the corresponding bias of the estimators.

\begin{figure}
\centering
\includegraphics[width=3in,angle=270]{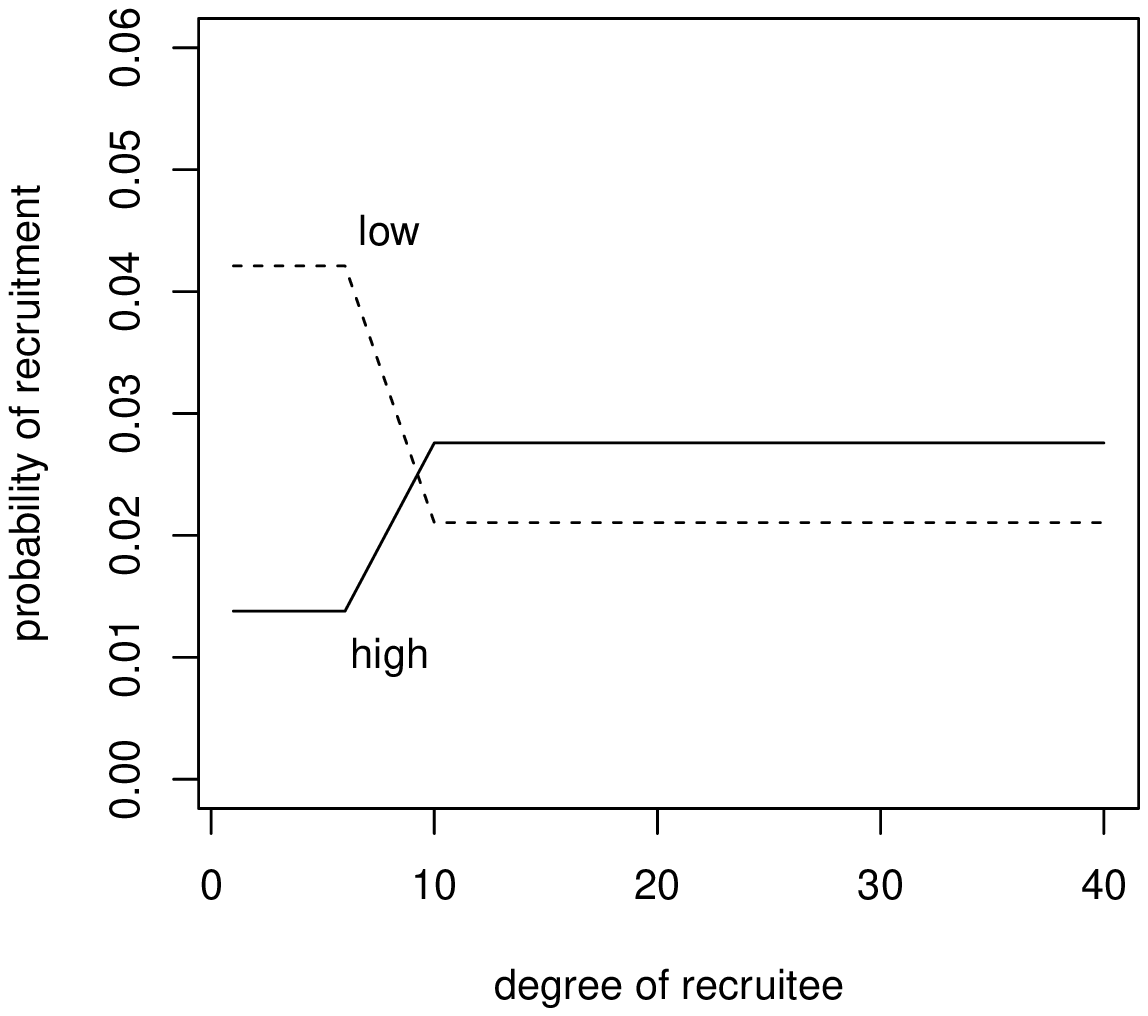}
\caption{Relative probabilities of recruitment for nodes, for different levels of differential recruitment between degree-group. The dashed line corresponds to preferential recruitment of low-degree nodes, and the solid line to preferential recruitment of high degree nodes. As with all simulations, the mean degree of all nodes is equal to 7.}
\label{fig:BetweenDegGrpProbs}
\end{figure}

\begin{figure}[h]
\centering
\includegraphics[angle=270,width=\textwidth]{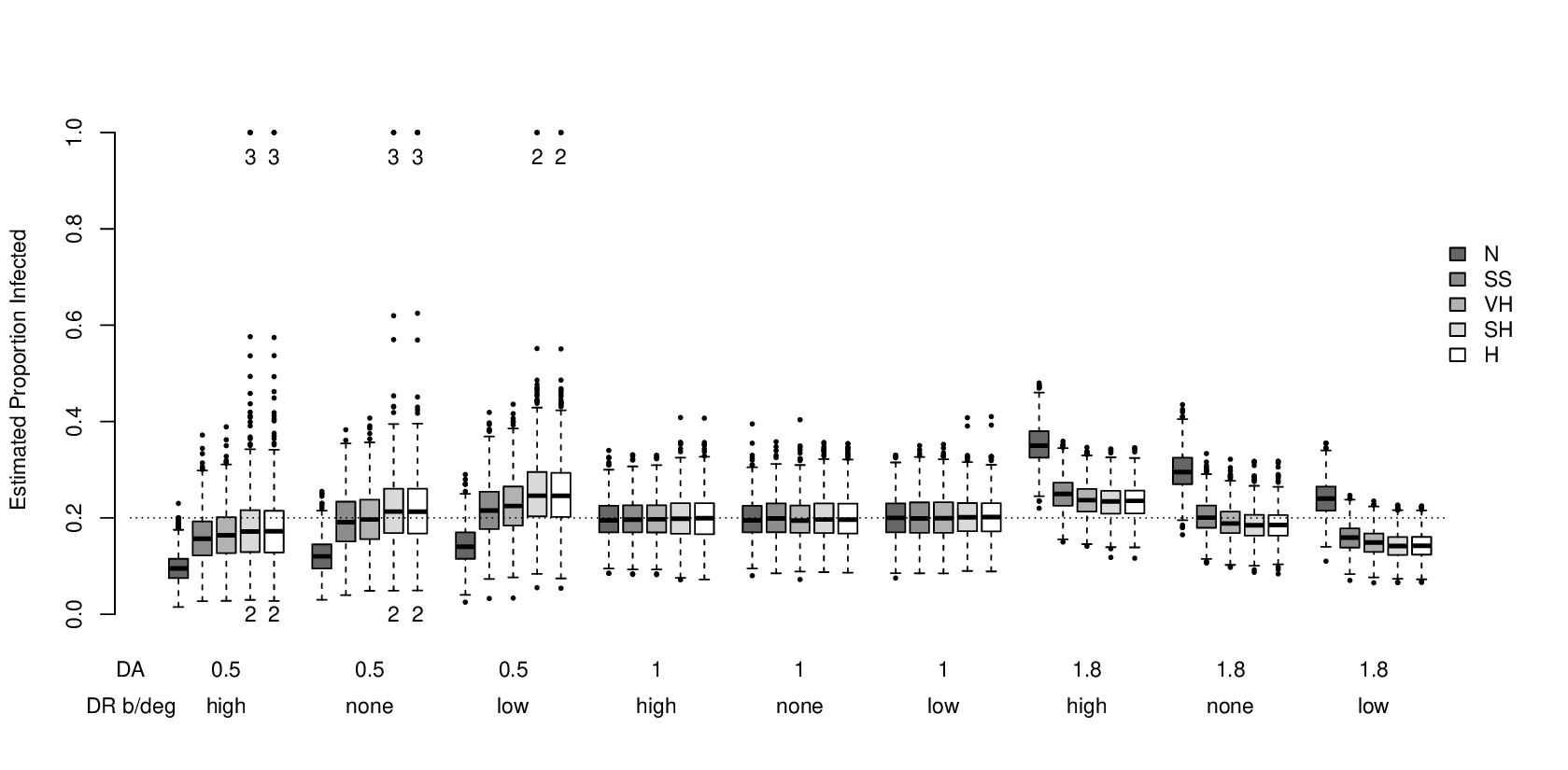}
\caption{Simulation results for varying levels of between-degree-group differential recruitment and differential activity. The labels indicate whether low (``low'') or high (``high'') degree nodes are preferentially recruited.}
\label{fig:DR.deg.b}
\end{figure}

For these simulations, we used three levels of differential recruitment: none, low degree nodes are preferentially recruited (``low''), and high-degree nodes are preferentially recruited (``high'').
For the latter two cases, the relative probabilities that a node of any degree will recruit a node of degree $d$ are illustrated in Figure \ref{fig:BetweenDegGrpProbs}.
Results of the simulations for the three levels of between-degree-group differential recruitment and three levels of differential activity are shown in Figure \ref{fig:DR.deg.b}.
From this figure we can see that when low-degree nodes are preferentially recruited this results in an increase in the mean of the estimators if differential activity is equal to $0.5$ and a decrease in mean of the estimators if differential activity is equal to $1.8$.
This is as we would expect based on the induced differential recruitment between infection groups.
There is no evidence that any of the estimators correct for this type of differential recruitment.

\section{Recruitment Effectiveness and Non-response}
\label{s:REandNR}

In this section we show that all estimators behave in a similar and predictable way in the presence of non-response.
We also explore the behaviour of the estimators when there is imperfect recruitment effectiveness, and find that the SH- and H- estimators behave differently to the other estimators considered.

We define recruitment effectiveness and non-response as follows:
\begin{definition}[Recruitment effectiveness] Denote by $R_g$ the probability that a coupon given to an individual in group $g$ is passed on to another candidate individual in the population, given that such an individual exists.
Then $R_g$ is the ``recruitment effectiveness'' of group $g$.
\end{definition}
\begin{definition}[Non-response] Denote by $V_g$ the probability that an individual in group $g$ reports to the study centre after having been given a coupon.
Then if $V_g \neq 1$ we say there exists non-response, and refer to $V_g$ as the ``response rate'' of group $g$.
\label{def:Vg}
\end{definition}
Recruitment effectiveness and response rate are closely linked.
If a coupon given to an individual is not returned, this could be either because that individual does not use the coupon to recruit anybody (imperfect recruitment effectiveness, $R_g < 1$), or because the person they recruit does not report back to the study centre (non-response, $V_g < 1$).
It is not always possible to tell reliably for which reason a coupon was not returned, nor to directly estimate recruitment effectiveness and non-response rates.

In this section we consider how imperfect recruitment effectiveness and non-response are likely to affect the estimators, and carry out a simulation study to test these arguments.

\subsection{Changes in the Selection Probabilities}
\label{ss:ChangesProbs}

Suppose that node $j \in g$ is given a coupon and there is probability $R_g < 1$ that he uses it to recruit another participant.
If $j$ does not pass on the coupon, then the children of $j$ miss out on a chance to be recruited.
Therefore, if $R_g < 1$ the children of nodes in group $g$ will have a lower probability of selection than if $R_g$ were equal to 1.
Hence in general one would expect that if $R_g < R_{g'}$, nodes who have many parents in group $g$ will have a lower probability of selection that nodes with many parents in group $g'$, everything else being equal.
If this effect is not balanced between infection groups, and in the presence of homophily, it is likely to result in biased estimates of $\PX$.
Because recruitment effectiveness does not affect the choice of recruit given that a coupon is passed on, changes in recruitment effectiveness would not be expected to affect the cross-group recruitment probabilities used in the SH- and H-estimators.

\subsection{Additional effect of Non-response}

Rather than changing coupon-passing probabilities, non-response affects which coupons are returned. 
Hence rather than reflecting the population proportion of infected individuals, $\PXhat$ will ``estimate'' the proportion of infected individuals among the population of respondents.
Suppose we denote the set of respondents in the population by $R$.
Then by definition \ref{def:Vg}, $V_g = N_{g\cap R}/N_g$ and so the population proportion of infected individuals can be written as
\begin{equation}
\nonumber
\frac{N_A}{N} = \frac{N_{A\cap R}}{N_{A\cap R} + N_{B\cap R}\frac{V_A}{V_B}}
\end{equation}
(derivation given in appendix \ref{app:nonresponse}).
If $V_A < V_B$, $N_A/N$ will be greater than $N_{A\cap R}/N_R$.
Thus if there are no other sources of bias, all the estimators will be biased downward.
If $V_A > V_B$, all estimators will be biased upward.

Similarly, non-response will affect the estimated cross-group recruitment probabilities.
This is because the estimates are based only on the recruitments to nodes who actually respond, which is a subset of all recruitments actually made. 
For example, if infected nodes have a lower response-rate than uninfected nodes, then the estimated probability of an uninfected node recruiting an infected node will be reduced compared to the case of perfect response.

\subsection{Results}

\subsubsection{Recruitment Effectiveness}

For our simulations, recruitment effectiveness by infection group is parameterised as a vector $(R_B,R_A)$, where $R_B$ and $R_A$ are the probability that a coupon is passed on given that a candidate for recruitment exists, for uninfected and infected nodes respectively.
We consider values $\{(0.6,0.9), {\bf (1,1)}, (0.9,0.6)\}$.

Recruitment effectiveness by degree group is parameterised in the form $(\alpha,\beta)$,  where nodes of degree 1--5 have relative probability $\alpha$ of recruiting, nodes of degree greater than 10 have relative probability $\beta$ of recruiting, and the relative probability for nodes of degree 6--10 increases linearly from $\alpha$ to $\beta$.  
Possible values are $\{(0.5,1),{\bf (1,1)},(1,0.5)\}$.

\begin{figure}[h]
\centering
\includegraphics[angle=270,width=\textwidth]{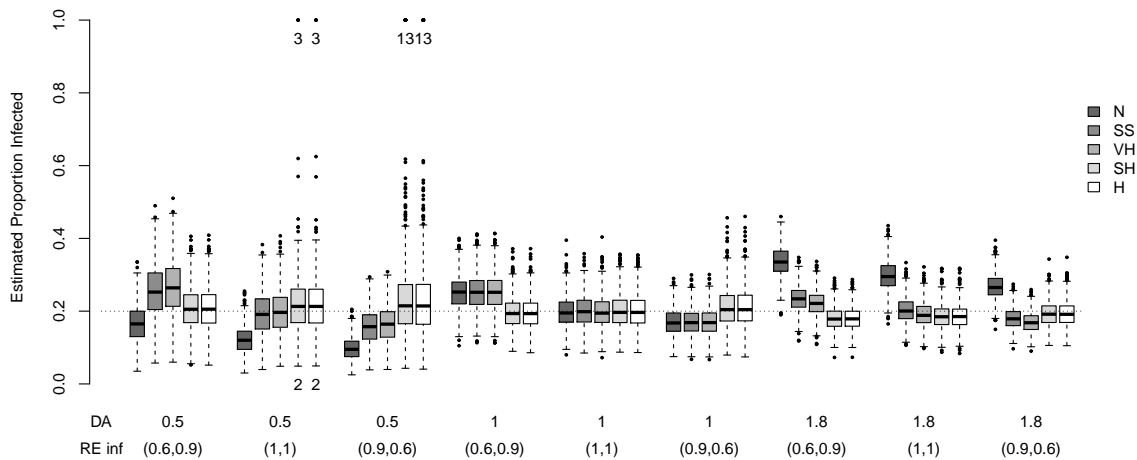}
\caption{Simulation results for varying levels of recruitment effectiveness by infection group and differential activity.}
\label{fig:RE.dis}
\end{figure}

\begin{figure}[h]
\centering
\includegraphics[angle=270,width=\textwidth]{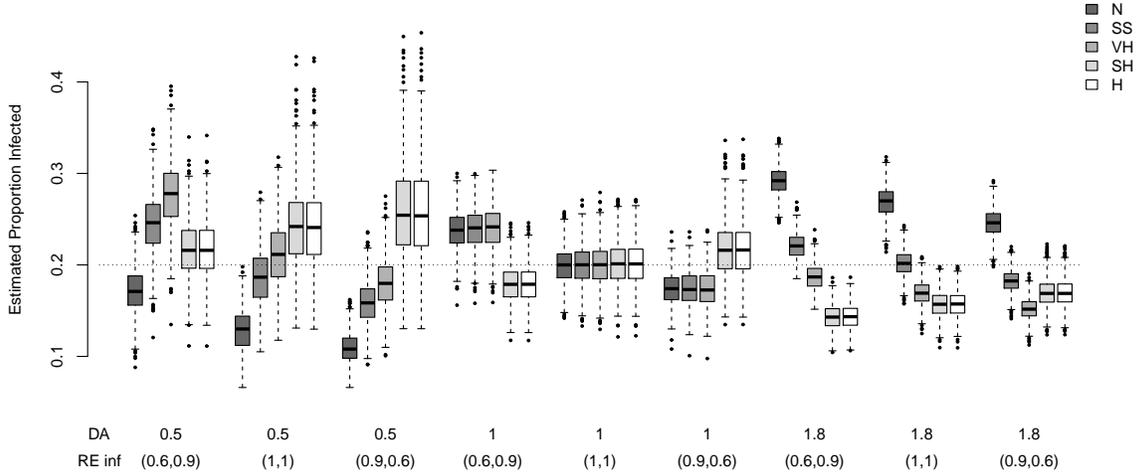}
\caption{Simulation results for varying levels of recruitment effectiveness by infection group and differential activity, for samples of size 500 rather than 200.}
\label{fig:RE.dis.500}
\end{figure}

Results of the simulations for the different levels of recruitment effectiveness by infection group are shown in Figure \ref{fig:RE.dis}.
Figure \ref{fig:RE.dis} shows that, for all levels of differential activity, as the recruitment effectiveness of uninfected nodes increases relative to the recruitment effectiveness of infected nodes, the sample proportion of infected nodes decreases.
This is as expected due to the population homophily.
The VH-estimates and SS- estimates also decrease as the sample proportion decreases, demonstrating that they are biased by different levels of recruitment effectiveness.
However, it is interesting to note that neither the SH- nor the H-estimates appear to be influenced to the same extent by recruitment effectiveness as the VH-, SS- or \Naive\ estimates.  
If there is no differential activity then the SH- and H- estimators appear to effectively control for different recruitment effectiveness within infection groups.
In the presence of differential activity there is a slight bias in the opposite direction to the \Naive\ estimator, but this tends to be small relative to the bias of the other estimators. 
We also ran the simulation for a sample size of 500.
In this case, shown in Figure \ref{fig:RE.dis.500}, the H- and SH- estimators have a substantial bias in the opposite direction to the \Naive\ and VH- estimators.
Thus the simulation results suggest that the SH- and H- estimates correct for bias due to differential recruitment in small sample fractions.
Again, there is no significant difference between the means of the SH- and H- estimators.\\

\begin{figure}[h]
\centering
\includegraphics[angle=270,width=\textwidth]{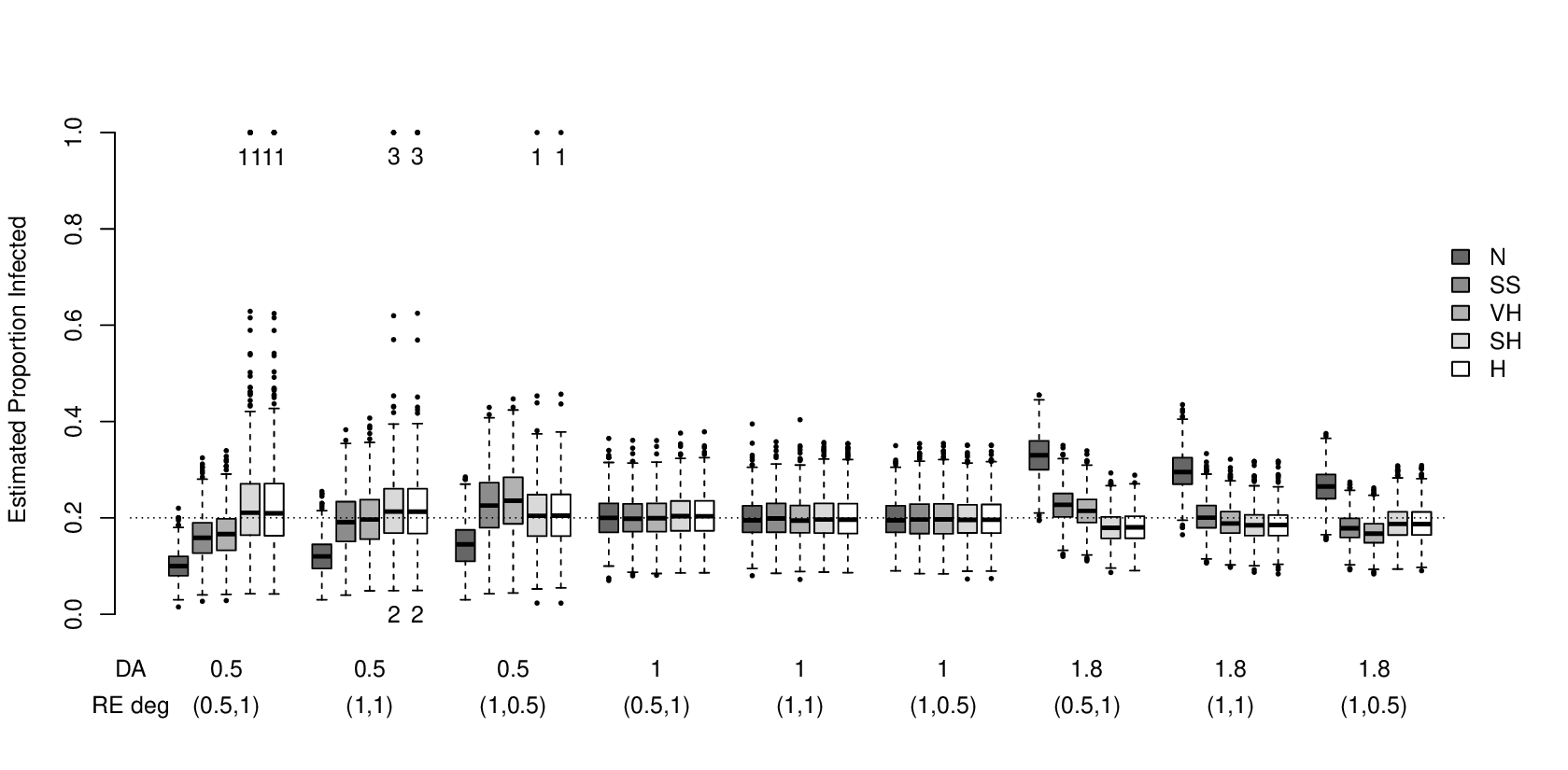}
\caption{Simulation results for varying levels of recruitment effectiveness by degree group and differential activity.}
\label{fig:RE.deg}
\end{figure}

In the case of different levels of recruitment effectiveness by degree-group, 
shown in Figure \ref{fig:RE.deg}, a similar effect can be seen.
In this case, if there is no differential activity then different levels of recruitment effectiveness by degree group does not bias the estimators of $\PX$, because the induced recruitment effectiveness by infection-group is the same for infected and uninfected groups.
For differential activity greater than 1, if low-degree nodes have higher recruitment rates than high-degree nodes (parameterisation $(1,0.5)$), this means that on average uninfected nodes will have higher recruitment rates than infected nodes.
Accordingly the sample proportion of infected individuals decreases as recruitment effectiveness by degree-group changes from $(0.5,1)$ to $(1,0.5)$ for differential activity equal to $1.8$, and the reverse is true if differential activity is equal to $0.5$.
Therefore, 
the patterns of bias are similar to the case of recruitment effectiveness by infection group.
The bias of the H- and SH- estimators increases accordingly for samples of size 500 (not shown).

The difference in mean between the SH- and H- estimators is statistically significant when recruitment effectiveness by degree-group is equal to $(0.5,1)$ and differential activity is $0.5$ or $1.8$. 
In both of these cases the mean of the H-estimates was closer to the true value than the mean of the SH-estimates.
However, from the simulation results of Figure \ref{fig:RE.deg}, this difference may not be large enough to be of practical significance.

\subsubsection{Non-response}

Non-response by infection group is parameterised as a vector $(V_B,V_A)$, where $V_B$ and $V_A$ are the probability that a node responds given that they have been recruited, for uninfected and infected nodes respectively.
Possible values are $\{(0.6,0.9), {\bf (1,1)}, (0.9,0.6)\}$.
Non-response by degree group is parameterised in the same way as recruitment effectiveness by degree group.
Possible values are $\{(0.5,1),{\bf (1,1)},(1,0.5)\}$.

\begin{figure}[h]
\centering
\includegraphics[angle=270,width=\textwidth]{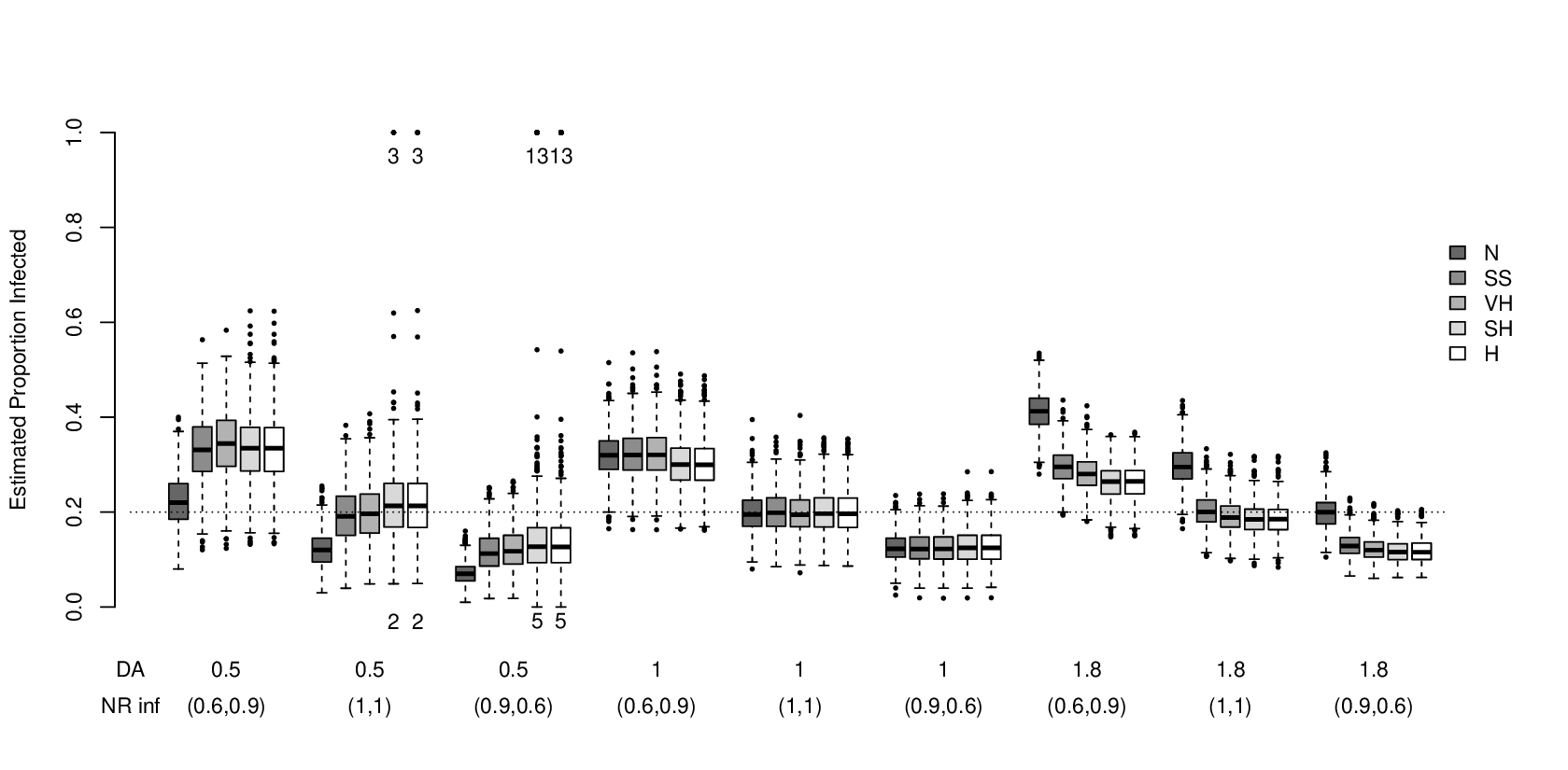}
\caption{Simulation results for varying levels of non-response by infection group and differential activity.}
\label{fig:NR.dis}
\end{figure}

\begin{figure}[h]
\centering
\includegraphics[angle=270,width=\textwidth]{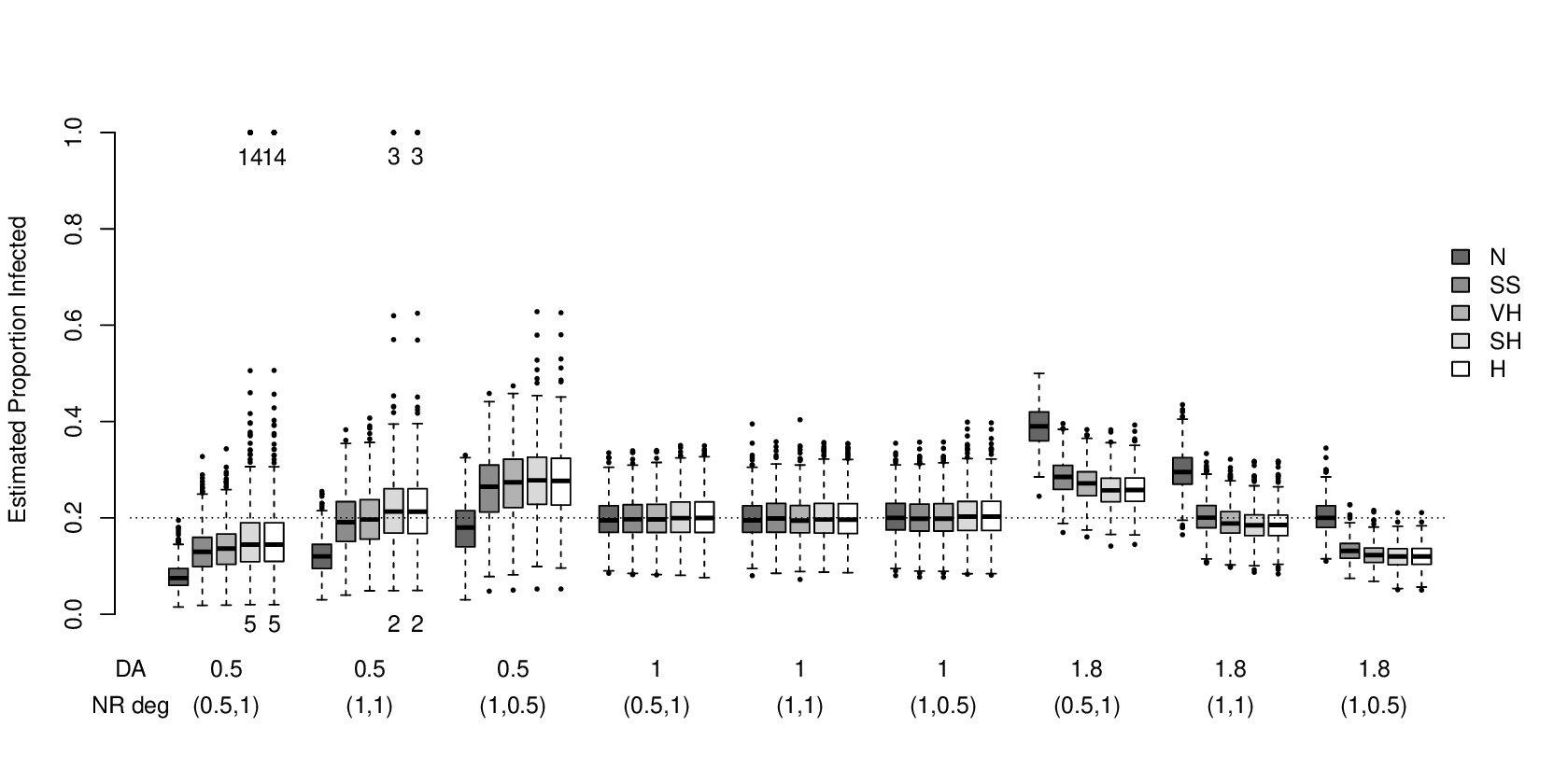}
\caption{Simulation results for varying levels of non-response by degree group and differential activity.}
\label{fig:NR.deg}
\end{figure}

Results of the simulations for the different levels of non-response by infection group and degree group are shown in Figures \ref{fig:NR.dis} and \ref{fig:NR.deg} respectively.
From Figure \ref{fig:NR.dis}, as the response-rate decreases for infected nodes relative to uninfected nodes, i.e. the parameterisation changes from $(0.6,0.9)$ to $(0.9,0.6)$, the sample proportion of infected nodes decreases, as expected.
Again, all estimators are biased by changes in the relative rates of non-response.
As the sample proportion of infected individuals decreases, so too do the \Naive, SS-, VH-, SH- and H- estimates.
There are no significant differences between the mean SH- and H- estimates.\\ 

For the case of non-response by degree-group, similarly to the case of recruitment effectiveness by degree-group, the results can be interpreted in terms of induced non-response by infection-group.
Viewed in this way, the results shown in Figure \ref{fig:NR.deg} are consistent with those of Figure \ref{fig:NR.dis}.
There are significant differences between the SH- and H- estimates when there exists both non-response by degree group and differential activity, but again these differences don't appear to be large. 
None of the estimators seem to control for different rates of non-response between infected and uninfected groups.

\section{Comparison of the Salganik-Heckathorn and Heckathorn Estimators}
\label{s:SHHdiff}

In Section \ref{ss:H} we hypothesised that the SH- and H- estimates would differ if the Markov chain on degree groups was started out of equilibrium.
Further, the more slowly the chain reaches equilibrium, the larger the expected difference between the SH- and H- estimates for a fixed sample size.

In the simulations presented in Sections \ref{s:DifferentialRecruitment} and \ref{s:REandNR} there were very few statistically significant differences between the mean SH- and H-estimates.
Those differences which were significant always occurred when there was differential activity and differences in recruitment or response behaviour between degree groups.
These are the cases where the equilibrium probabilities of the Markov chain on degree groups are likely to be furthest from the degree distribution of the seeds.
However, in all simulations the differences between the mean (and median) estimates of the H- and SH- estimators were unlikely to be large enough to be of practical concern (all mean differences were less than $0.001$), and the variances of the estimates were also nearly identical.

\citet[pp. 185]{H2007} admits that traditional RDS designs are likely to result in only ``modest'' differences between the two estimators.
However, he theorises that
\begin{quote}
``\ldots changes in research design that induce an association between degree and opportunities to recruit would produce much larger potential effects.''
\end{quote}
A design that gave respondents with higher reported degree more coupons than respondents with low reported degree would have such an effect.
Such a design can be viewed as equivalent to one where all nodes are given the same number of coupons but nodes of lower degree have much lower recruitment effectiveness.
This case was considered in Section \ref{s:REandNR}, Figure \ref{fig:RE.deg}, and again produced only very small differences between the H- and SH- estimates.
Similarly, a design where infected nodes are given more coupons than uninfected nodes is equivalent to the case we considered in Figure \ref{fig:RE.dis}, where uninfected nodes have lower recruitment effectiveness than infected nodes.
Again, only very small differences between the H- and SH- estimates were observed.
We also re-ran the simulations using four coupons rather than two, but this did not produce any significant changes in the results.  

\begin{figure}[h]
\centering
\includegraphics[angle=270,width=\textwidth]{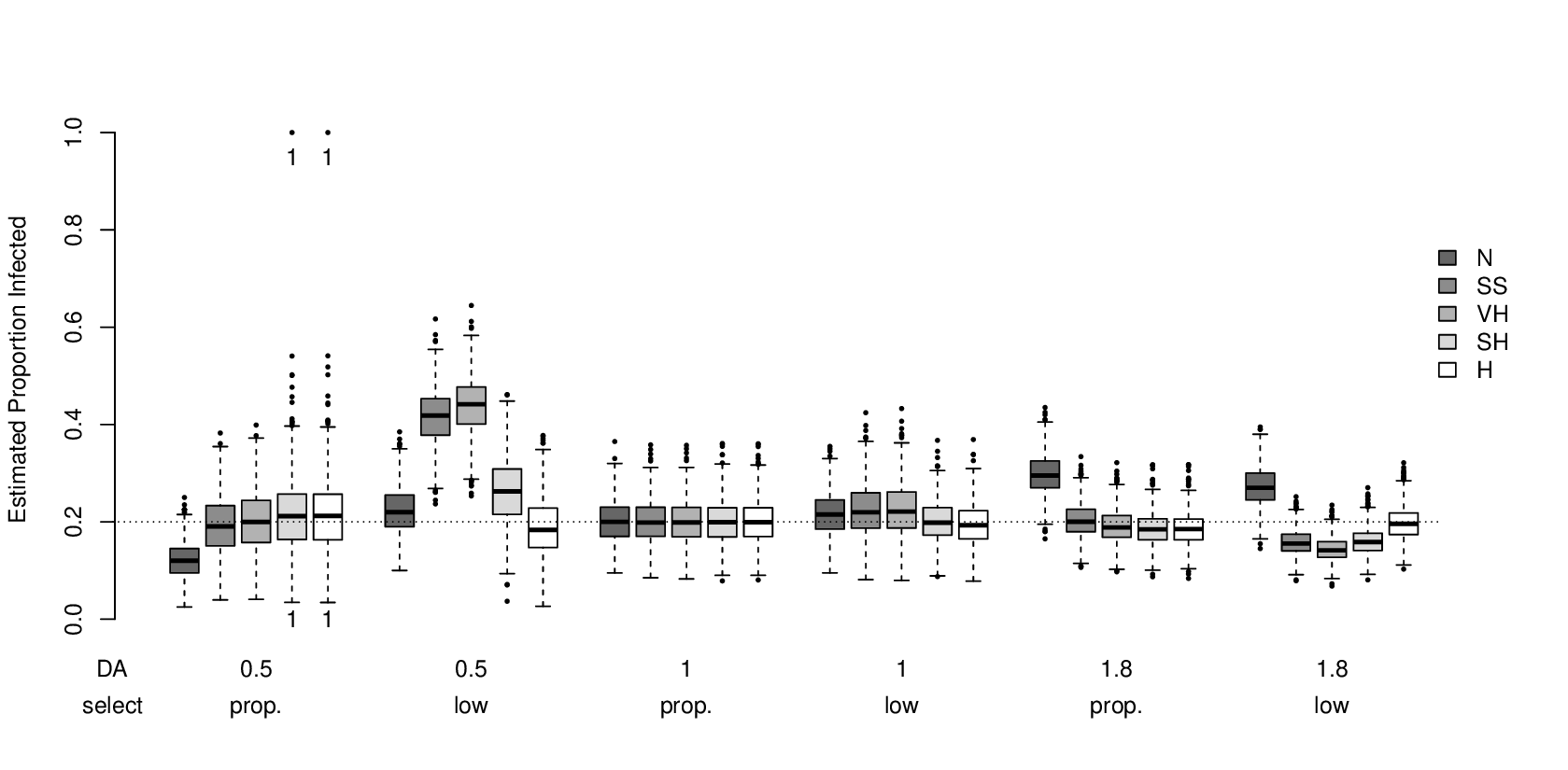}
\caption{Box-plots of the estimates for different methods of selecting seeds. Seeds are either selected with probability proportional to degree or uniformly at random from the twenty nodes of lowest degree.}
\label{fig:select}
\end{figure}

To try and induce larger differences between the estimates we looked at the effect of increasing homophily (to slow convergence), preferentially selecting seeds from nodes of low or high degree (to start the chain out of equilibrium), and more extreme differences in recruitment effectiveness.
The largest differences between the SH- and H- estimates arose when the seeds were selected from nodes of low degree.
Compared to the effect of seed selection, increasing differential activity or homophily had little effect on the difference between the estimates.
Simulation results for the case that ten seeds are chosen uniformly at random from the twenty nodes of lowest degree are shown in Figure \ref{fig:select}.
The observable differences between the H- and SH- estimates when seeds are selected from low-degree nodes are statistically significant.
It is also very interesting to note that in these cases the H-estimator appears to correct somewhat for the bias introduced by the seed selection mechanism.

\begin{figure}[h]
\centering
\includegraphics[angle=270,width=4.5in]{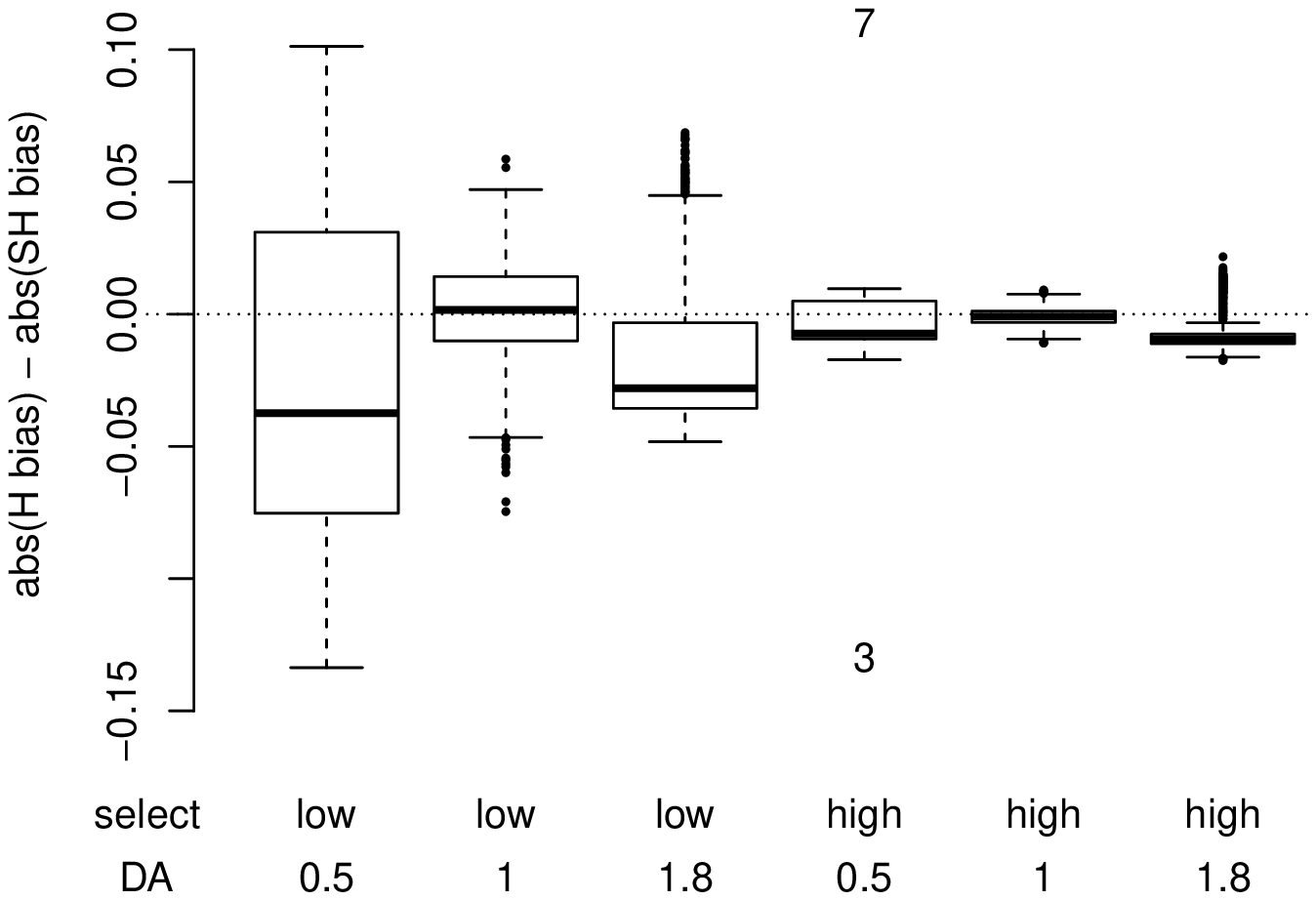}
\caption{Box-plots of the difference in absolute deviations from $0.2$ of the estimates returned by the SH- and H- estimators when seeds are chosen either from among the nodes of lowest or highest degree.}
\label{fig:select.lowhigh}
\end{figure}

To investigate further, we considered the differences in absolute deviations from $0.2$ of the H- and SH- estimates.
Figure \ref{fig:select.lowhigh} shows that on average the H- estimate is closer to the true value of $0.2$ than the SH- estimate when seeds are chosen from nodes of either high or low degree.
This observation is robust to changes in sample-size.
The simulation results therefore suggest that the H- estimator corrects for bias due to selecting seeds with degrees not representative of the theoretical equilibrium distribution on degree groups.
If there are no other sources of bias, the H- estimator is likely to be less biased than the SH- estimator in this case.
However, it should be noted that the seed selection scenarios which produced the observed differences in behaviour were quite extreme, and are unlikely to occur unless by design.

We also considered how the SH- and H- estimators behave when seeds are selected from nodes of low degree and, in addition, there exists differential recruitment, unbalanced recruitment effectiveness or non-response.
In this case the biases due to seed selection and recruitment effectiveness or non-response behaved in an approximately additive fasion.
Hence depending on the components of the additive bias, either the SH- or H- estimates could have the lower total bias.

In practice, a scenario as extreme as that illustrated in Figure \ref{fig:select} is unlikely to arise.
More realistically, it could happen that there exists differential activity and that seeds are selected only from infected individuals. 
If infected nodes have a lower or higher average degree than uninfected nodes, then this will induce the same effect as considered above.
We ran the simulations for this case and, although the difference between the mean of the H- and SH- estimators was statistically significant for the cases with differential activity and seeds chosen only from infected nodes (and only those cases), the size of the differences (approximately $0.001$) is unlikely to be of practical concern.
This indicates that it is unlikely that these conditions will result in large differences between H- and SH- estimates in traditional respondent driven sampling designs.

\begin{sidewaystable}[h]
\caption{Summary of conditions under which tested respondent behaviours induce bias}
\centering
\begin{tabular}{rlccccc}
 & & & & Estimator & & \\
Condition  & Description & N & SS & VH & SH & H  \\
\hline \hline
{\bf Differential Recruitment} & Non-random coupon-passing among alters & \\
~~~~~~ Within-Group & Different rates of recruitment to own group &\\
~~~~~~ ~~~~~~ Infection & To own infection group & \checkmark & \checkmark & \checkmark & \checkmark & \checkmark\\
~~~~~~ ~~~~~~ Degree & To own degree group & $\dagger$ &&&&\\
\hline
~~~~~~ Between-Group & One group preferentially recruited & \\
~~~~~~ ~~~~~~ Infection & Infection group preferred & \checkmark & \checkmark & \checkmark & \checkmark & \checkmark\\
~~~~~~ ~~~~~~ Degree & Degree group preferred & $\dagger$ & $\dagger$ & $\dagger$ & $\dagger$ & $\dagger$\\
\hline \hline
{\bf Recruitment Effectiveness} & Differential rates of coupon-passing & \\
~~~~~~ Infection Group & Infection group passes more coupons & \checkmark & \checkmark & \checkmark & \checkmark$^*$ & \checkmark$^*$ \\
~~~~~~ Degree Group & Degree group passes more coupons & $\dagger$ & $\dagger$ & $\dagger$ & $\dagger^*$ & $\dagger^*$ \\
\hline \hline
{\bf Non-Response} & Differential rates of coupon return &\\
 ~~~~~~ Infection Group & Infection group returns more coupons & \checkmark & \checkmark & \checkmark & \checkmark & \checkmark\\
~~~~~~ Degree Group & Degree group returns more coupons & $\dagger$ & $\dagger$ & $\dagger$ & $\dagger$ & $\dagger$\\
\hline \hline
{\bf Low-degree Seeds} & Seeds selected from low degree nodes & \checkmark & \checkmark & \checkmark & $\dagger$ & \\
\hline \hline
{\bf Key} & \\
 ~~~~~~ \checkmark & Estimator shows bias regardless of differential activity\\
 ~~~~~~ $\dagger$ & Bias only in the presence of differential activity\\
 ~~~~~~ $^*$ & Bias only with large sample fraction \\ 
\end{tabular}
\label{tab:summary}
\end{sidewaystable}

\section{Discussion}
\label{s:Conclusion}

In this paper we compare the performance of several respondent driven sampling estimators under conditions of differential recruitment, recruitment effectivness and non-response.  Our specific findings are summarized in Table \ref{tab:summary}.
Of particular importance is the performance of the estimator introduced in \cite{H2007}, which claims to adjust for differential recruitment and recruitment effectiveness.

We did not find any evidence that the H-estimator adjusts for differential recruitment or non-response.
However, in the case of recruitment effectiveness, the resulting bias of the H-estimator and the SH- estimator is in the opposite direction to that of the \Naive, SS- and VH- estimators, and the size of this bias depends on the sampling fraction (see Figures \ref{fig:RE.dis} and \ref{fig:RE.dis.500}).
For the simulations with sample size 200, this resulted in the SH- and H- estimators effectively controlling for the bias due to imperfect recruitment effectiveness. 

When seeds were randomly selected with probability proportional to degree, under all conditions of differential recruitment, recruitment effectiveness and non-response the estimates of the H-estimator were almost identical to those of the SH-estimator.
In Sections \ref{s:DifferentialRecruitment} (differential recruitment) and \ref{s:REandNR} (recruitment effectiveness and non-response), we found statistically significant differences between these two estimators in only 10 of 69 simulation cases (sample size 200).  
The magnitudes of the differences in these cases were small enough to be negligible for practical purposes.  

This led us to explore, in Section \ref{s:SHHdiff}, other conditions which might lead to differences between the H- and SH- estimates.   
Here, we found that we can induce differences between the two estimators when the following two conditions both hold:
\begin{enumerate}
\item The seeds are selected with probability far from the theoretical equilibrium probabilities of the Markov chain on degree groups; and
\item The degree distribution of infected and uninfected nodes is not the same (for example if there is differential activity).
\end{enumerate}
Under these conditions the H-estimator was significantly less biased than the SH-estimator, indicating that the H-estimator may correct for the bias introduced by this type of seed selection.
Specifically, we found the largest differences between the estimates of the H- and SH- estimators when seeds were selected from among the nodes of lowest degree.
Smaller, but substantial differences were also apparent when selecting seeds from among nodes of the highest degree.
It is possible this scenario may arise by design, where study designers try to recruit well-connected population members as seeds.  

If there are no sources of bias other than unbalanced selection of seeds by degree, then the H- estimator is likely to have a smaller bias than the other estimators considered in this paper.
However, if other sources of bias are present (which will almost always be the case), then these will interact to mean that the absolute bias of the H- estimator may not be the smallest of the estimators considered.

All estimators were subject to bias induced by differential recruitment, recruitment effectiveness and non-repsonse.  
It is interesting, however, that when the differential recruitment and recruitment effectiveness acted only on degree groups, bias was only present in the estimators when the infection groups had different degree distributions (differential activity $\neq 1$).  
Without this condition, over or under representing degree groups and their sampling patterns in the estimators does not differentially affect the two groups.  
Across these simulations, for random selection of seeds the H- and SH- estimators tended to perform similarly in each case. 
The \Naive\ estimator (sample mean) tended to exhibit the largest bias, whereas the SS-, VH-, SH- and H- estimators showed reduced bias due to adjusting for differential activity \citep{GH2009}.
For all cases the SS- estimates fell between the \Naive\ and VH- estimates.
In general, apart from the conditions of seed selection mentioned above, none of the SS-, VH-, SH- or H- estimators were consistently less biased than any other.
They also had similar variances, although variance of the VH- and SS- estimators appeared to be slightly less in many cases, and the estimates returned by the SS- and VH- estimators were not subject to the same instability as the SH- and H- estimates.

Overall, our findings, indicate that no estimator consistently out-performs the others under all conditions.  
We find that with small sample fractions, the SH- and H- estimators correct for differential recruitment effectiveness across groups, which induces bias in the other estimators in the presence of homophily.  
This is an advantage, and therefore these estimators should be used whenever the sample fraction is small and differential recruitment effectiveness is suspected.  
However, these estimators also are also subject to more instability, and have larger variance and greater bias than the VH- or SS- estimators under many other sample conditions, both in this paper, and as described in \cite{GH2009}.  
This suggests that absent differential recruitment effectiveness, the VH- or SS- estimators are to be preferred.  
In the case of large sample fractions, the SS- estimator is clearly to be preferred \citep{G2010}.  
Our results do not provide evidence that the H-estimator should be preferred over the SH-estimator.  
Although we found differences between the H- and SH- estimators in some extreme cases, we suspect these cases are rare enough so as to not justify the additional complexity of the H-estimator.
The lack of clear preference among these estimators highlights the need for both new estimators, and also for new diagnostic tests aimed at identifying circumstances in which each estimator is to be preferred.  

It is also important to keep in mind the limitations of this study.
It was not possible to consider all possible combinations and levels of the parameterisations of the effects studied in this paper.  
We have tried to focus on a range of simulation conditions and parameter values most likely to highlight contributions of the H-estimator, and considered variations of many parameters at once to draw out potential interaction effects.  
Nevertheless, there may well be other population or sampling parameters whose levels will impact the results of our study.  
While we have tried to include any confounders we felt might affect our qualitative findings, any effect sizes should be understood as specific to the simulation conditions under study.\\

\appendix

\section{Derivation of Hansen-Hurwitz Estimators}
\label{app:HHest}

Consider the Hansen-Hurwitz estimator of $\Pi_A$,
\begin{eqnarray}
\PiXhat &=& 
\sum_{i \in \sX}\pi_i^{-1}\pi_i \nonumber \\
 &=& {\nX}. \nonumber
\end{eqnarray}
Thus, taking the ratio of Hansen-Hurwitz estimators for $\Pi_A$ and $\Pi_A + \Pi_B$ shows that $\nX/(\nX+\nY)$ is a generalised Hansen-Hurwitz estimator of $\Pi_A/(\Pi_A + \Pi_B)$.

\section{Non-response}
\label{app:nonresponse}

By definition \ref{def:Vg}, $V_g = N_{g\cap R}/N_g$ and so the population proportion of infected individuals can be written as
\begin{eqnarray}
\frac{N_A}{N} &=& \frac{V_BV_AN_A}{V_BV_AN_A + V_AV_BN_B} \nonumber \\
 &=& \frac{V_BN_{A\cap R}}{V_BN_{A\cap R} + V_AN_{B\cap R}}\nonumber \\
 &=& \frac{N_{A\cap R}}{N_{A\cap R} + N_{B\cap R}\frac{V_A}{V_B}}\nonumber
\end{eqnarray}
as claimed.

\bibliography{../../../rdsbib}

\begin{thebibliography}{}

\bibitem[\protect\citeauthoryear{Abdul-Quader, Heckathorn, McKnight, Bramson,
  C.~Nemeth, and Jarlais}{Abdul-Quader et~al.}{2006}]{AQ2006}
Abdul-Quader, A.~S., D.~D. Heckathorn, C.~McKnight, H.~Bramson, K.~G.
  C.~Nemeth, and D.~C.~D. Jarlais (2006).
\newblock Effectiveness of respondent-driven sampling for recruiting drug users
  in {N}ew {Y}ork {C}ity: {F}indings from a pilot study.
\newblock {\em Journal of Urban Health\/}~{\em 83}, 459--476.

\bibitem[\protect\citeauthoryear{Gile}{Gile}{2010}]{G2010}
Gile, K.~J. (2010).
\newblock Improved inference for respondent-driven sampling data with
  application to {HIV} prevalence estimation.
\newblock {\em Journal of the American Statistical Association\/}.
\newblock To appear. \url{http://arxiv.org/abs/1006.4837v1}.

\bibitem[\protect\citeauthoryear{Gile and Handcock}{Gile and
  Handcock}{2010}]{GH2009}
Gile, K.~J. and M.~S. Handcock (2010).
\newblock Respondent-driven sampling: An assessment of current methodology.
\newblock {\em Sociological Methodology\/}~{\em 40\/}(1), 285--327.

\bibitem[\protect\citeauthoryear{Goel and Salganik}{Goel and
  Salganik}{2009}]{GS2009}
Goel, S. and M.~J. Salganik (2009).
\newblock Respondent-driven sampling as markov chain monte carlo.
\newblock {\em Statistics in Medicine\/}~{\em 28\/}(17), 2202--2229.

\bibitem[\protect\citeauthoryear{Handcock, Gile, and Neely}{Handcock
  et~al.}{2009}]{rdspackage}
Handcock, M.~S., K.~J. Gile, and W.~W. Neely (2009).
\newblock {\em \pkg{RDS}: R Functions for Respondent-Driven Sampling}.
\newblock Seattle, WA: Hidden Population Methods Research Group
  \url{http://hpmrg.org/}.
\newblock \proglang{R}~package version~0.10.

\bibitem[\protect\citeauthoryear{Handcock, Hunter, Butts, Goodreau, and
  Morris}{Handcock et~al.}{2003}]{statnet}
Handcock, M.~S., D.~R. Hunter, C.~T. Butts, S.~M. Goodreau, and M.~Morris
  (2003).
\newblock {\em \pkg{statnet}: Software Tools for the Statistical Modeling of
  Network Data}.
\newblock Seattle, WA: Statnet Project \url{http://statnetproject.org/}.
\newblock \proglang{R}~package version~2.0.

\bibitem[\protect\citeauthoryear{Handcock, Hunter, Butts, Goodreau, and
  Morris}{Handcock et~al.}{2008}]{statnetjss}
Handcock, M.~S., D.~R. Hunter, C.~T. Butts, S.~M. Goodreau, and M.~Morris
  (2008).
\newblock \pkg{statnet}: Software tools for the representation, visualization,
  analysis and simulation of social network data.
\newblock {\em Journal of Statistical Software\/}~{\em 24\/}(1).
\newblock \url{http://www.jstatsoft.org/v24/i01/}.

\bibitem[\protect\citeauthoryear{Hansen and Hurwitz}{Hansen and
  Hurwitz}{1943}]{HH1943}
Hansen, M. and W.~Hurwitz (1943).
\newblock On the theory of sampling from finite populations.
\newblock {\em Annals of Mathematical Statistics\/}~{\em 14}, 333--362.

\bibitem[\protect\citeauthoryear{Heckathorn}{Heckathorn}{1997}]{H1997}
Heckathorn, D.~D. (1997).
\newblock Respondent-driven sampling: A new approach to the study of hidden
  populations.
\newblock {\em Social Problems\/}~{\em 44}, 174--199.

\bibitem[\protect\citeauthoryear{Heckathorn}{Heckathorn}{2007}]{H2007}
Heckathorn, D.~D. (2007).
\newblock Extensions of respondent-driven sampling: Analyzing continuous
  variables and controlling for differential recruitment.
\newblock {\em Sociological Methodology\/}~{\em 37\/}(1), 151--207.

\bibitem[\protect\citeauthoryear{Horvitz and Thompson}{Horvitz and
  Thompson}{1952}]{HT1952}
Horvitz, D.~G. and D.~J. Thompson (1952).
\newblock A generalization of sampling without replacement from a finite
  universe.
\newblock {\em Journal of the American Statistical Association\/}~{\em 47},
  663--685.

\bibitem[\protect\citeauthoryear{Neely}{Neely}{2009}]{N2009}
Neely, W.~W. (2009).
\newblock Statistical theory \& respondent-driven sampling (under review).

\bibitem[\protect\citeauthoryear{Salganik and Heckathorn}{Salganik and
  Heckathorn}{2004}]{SH2004}
Salganik, M.~J. and D.~D. Heckathorn (2004).
\newblock Sampling and estimation in hidden populations using respondent-driven
  sampling.
\newblock {\em Sociological Methodology\/}~{\em 34}, 193--239.

\bibitem[\protect\citeauthoryear{Volz and Heckathorn}{Volz and
  Heckathorn}{2008}]{VH2008}
Volz, E. and D.~D. Heckathorn (2008).
\newblock Probability based estimation theory for respondent driven sampling.
\newblock {\em Journal of Official Statistics\/}~{\em 24\/}(1), 79--97.

\bibitem[\protect\citeauthoryear{Volz, Wejnert, Degani, and Heckathorn}{Volz
  et~al.}{2007}]{RDSAT}
Volz, E., C.~Wejnert, I.~Degani, and D.~D. Heckathorn (2007).
\newblock Respondent-driven sampling analysis tool (rdsat) version 5.6.
\newblock \url{www.respondentdrivensampling.org}.

\end{thebibliography}

\end{document}